\documentclass[twocolumn,longbibliography]{quantumarticle-customized}
\usepackage{amsmath}
\usepackage{amsfonts}
\usepackage{graphicx}
\usepackage[pdfpagelabels,pdftex,bookmarks,breaklinks]{hyperref}
\usepackage[all]{hypcap}
\hypersetup{colorlinks,citecolor=blue,urlcolor=blue,linkcolor=blue}

\title{Factoring with $n+2$ clean qubits and $n-1$ dirty qubits}
\author{Craig Gidney}
\affiliation{Google, Santa Barbara, CA 93117, USA}
\email{craiggidney@google.com}

\begin{document}
\maketitle

\begin{abstract}
We present reversible classical circuits for performing various arithmetic operations aided by dirty ancillae (i.e. extra bits/qubits in an unknown state that must be restored before the circuit ends).
We improve the number of clean qubits needed to factor an $n$-bit number with Shor's algorithm \cite{Shor1999} from $1.5n + O(1)$ \cite{zalka2006} to $n+2$, assisted by $n-1$ dirty qubits, without increasing the asymptotic size or depth of the circuit.
\end{abstract}

\section{Introduction} \label{sec:introduction}

When constructing quantum circuits, or classical reversible circuits, an important resource is the number of available ancillae.
An ancilla is an extra bit or qubit that is available for use by a circuit as temporary workspace.
Ancillae may be initialized to a known computational basis state (``clean bit"), or be given to the circuit in an unknown and potentially entangled state that must be restored before the circuit finishes (``dirty bit").
Clean bits are more valuable, allowing for simpler and more compact circuit constructions, but dirty bits are more plentiful, since any temporarily unused bit is a borrowable dirty bit.

One part of a circuit can borrow dirty bits from another part of the same circuit, so circuit constructions that require only dirty bits are easier to apply under tight space constraints, or on circuit topologies where other ancillae are too far away to be acquired quickly.
When attempting to reduce the number of bits or qubits required by a circuit, replacing constructions that use clean ancillae with ones that use dirty ancillae is a useful intermediate goal.

It is important to note that, pragmatically speaking, it is far more important to, for example, achieve low T gate counts under plausible architectural constraints than to reduce the number of clean qubits required by a circuit.
Our goal in this paper is not to come up with an implementation of Shor's algorithm optimized in the ways that matter for plausible future quantum computer architectures.
Our goal is to explore the consequences of picking a metric, trying to optimize it, and seeing what ideas fall out.
The circuit constructions we present will not be optimized to achieve good constant factors on circuit depth or gate count (though their asymptotics are fine).
And they will ignore machine architecture; they assume all-to-all connectivity between qubits.
And the {\em total} number of qubits we use ($2n+1$) is higher than previous work.
But interesting ideas don't always come from thinking directly about a problem; there is much to be found by placing artificial goals and seeing where they lead.
That exploration is our intent.
In this paper, we reduce the number of clean qubits required to factor an $n$-bit number with Shor's algorithm from $1.5n + O(1)$ clean qubits \cite{zalka2006} to $n+2$ clean qubits assisted by $n-1$ dirty qubits.
We do so without increasing the asymptotic circuit depth or size.

Our paper is structured as follows.
\autoref{sec:introduction} introduces and describes the conventions our circuit constructions and circuit diagrams will follow.
In \autoref{sec:construct} we describe all the circuit constructions we use to reduce the period finding step of Shor's algorithm into constant-sized gates, while tracking the number of required dirty ancillae.
Then, in \autoref{sec:costs}, we discuss the novelty and comparative costs of the presented circuit constructions.
Finally, \autoref{sec:conclusion} concludes with a discussion of possible future improvements.

All constructions in this paper use a two's-complement representation of integers.
When a result would exceed the size of a register, it wraps (i.e. all the non-modular arithmetic constructions we discuss actually perform arithmetic modulo $2^n$, where $n$ is the size of the target register).

All diagrams order qubits from least significant bit (LSB) at the top to most significant bit (MSB) at the bottom.

All our circuit diagrams annotate operations with the number of clean and/or dirty ancillae they need.
For example, a dashed line from an operation down to a triangle inscribed with ``3 dirty" means that the operation needs 3 dirty ancilla.
If there are 3 unused wires that the operation can borrow, then the triangle will be green and a note of ``(satisfied)" will be written underneath.
If there are not enough unused wires shown in the diagram, the triangle will be yellow.

The ancillae counts shown in diagrams and discussed in the text are not optimal; they are entirely based on what the constructions we discuss in this paper achieve.
Nearly all the circuit constructions we present are classical reversible circuits (i.e. they do not use any quantum operations), so tricks such as the ancilla bootstrap shown in \autoref{fig:bootstrap-ancilla} could cut the ancilla count in several places.
However, none of these improvements decrease the ancillae required at any crucial bottleneck, and so they do not reduce the number of qubits required by our overall construction of Shor's algorithm.

To avoid ambiguity, circuit diagrams will divide multi-register operations into separate parts.
For example, for an addition operation that adds a register $x$ into a register $y$ (i.e. performs $|x\rangle |y\rangle \rightarrow |x\rangle |y + x \pmod{2^n} \rangle$), we will place a light-gray box with the text ``Input A" over the wires corresponding to $x$ and a white box with the text ``+A" over the wires corresponding to $y$.
The ``A" symbol refers to the value of the input register, in the computational basis, at the time of the operation.
Diagrams will often use a sequence of operations each with their own ``Input A" box specific to that operation.
This reuse of ``A" does not indicate any relation between those input values.
The use of ``A" is merely a convention for indicating how the input-indicating box and the effect-indicating box are supposed to combine to form a single operation.

In the case of operations parametrized by compile-time constants, diagrams will use the letters ``K" and/or ``R".
For example, a box showing ``+K (mod R)" refers to the modular-offset operation $|x\rangle \rightarrow |x+K \pmod{R}\rangle$.
We use $R$ to refer to the modulus in modular-arithmetic operations, and to the value that is factored by Shor's algorithm.
In several constructions we assume that $R$ is odd.
Removing factors of 2 from a factoring problem is trivial, so this assumption does not decrease the generality of the overall construction.

Because adding a constant into a register is more expensive than adding one register into another, we avoid ambiguity between the two operations by always referring to the operation that adds a constant as an ``offset operation".
We will refer to the operation that adds one register into another as just ``addition" or, if needed for clarity, ``enregistered addition".

When an operation is controlled by a wire bundle, it is conditioned on all wires in the bundle.
When a single-qubit gate is applied to a wire bundle, the gate applies separately to every wire.
For example, when a NOT gate controlled by one wire bundle is applied to another wire bundle, every wire in the latter is toggled when and only when every wire in the former is on.

For completeness, even when the period finding reduction doesn't require a controlled version of an operation, we nevertheless provide a controllable construction that scales linearly with the number of controls.

For brevity, we take for granted that each circuit's inverse operation is decomposed into the same operations as the original operation but run in reverse order and with each sub-operation inverted.
A decrement is a reversed increment, a multiply-add is a reversed multiply-subtract, a modular division is a reversed modular multiplication, a modular halving is a reversed modular double, and so forth.

We do not attempt to define or implement reasonable behavior for modular arithmetic circuits applied to out of range values (ones equal to or larger than the modulus).
This {\em includes input values}.
For example, if our modular addition construction ($x \mathrel{{+}{=}} y \pmod{R}$) is applied to a $y \geq R$, then we don't guarantee that $y \mod R$ will be added into $x$, that the operation will commute with other additions, that optimizations won't change the behavior, or that anything sensible at all will happen.
If the precondition $y < R$ is violated, what occurs is undefined behavior \cite{c11}.

\section{Constructions} \label{sec:construct}

\begin{figure}
  \centering
  \includegraphics[height=17.3cm]{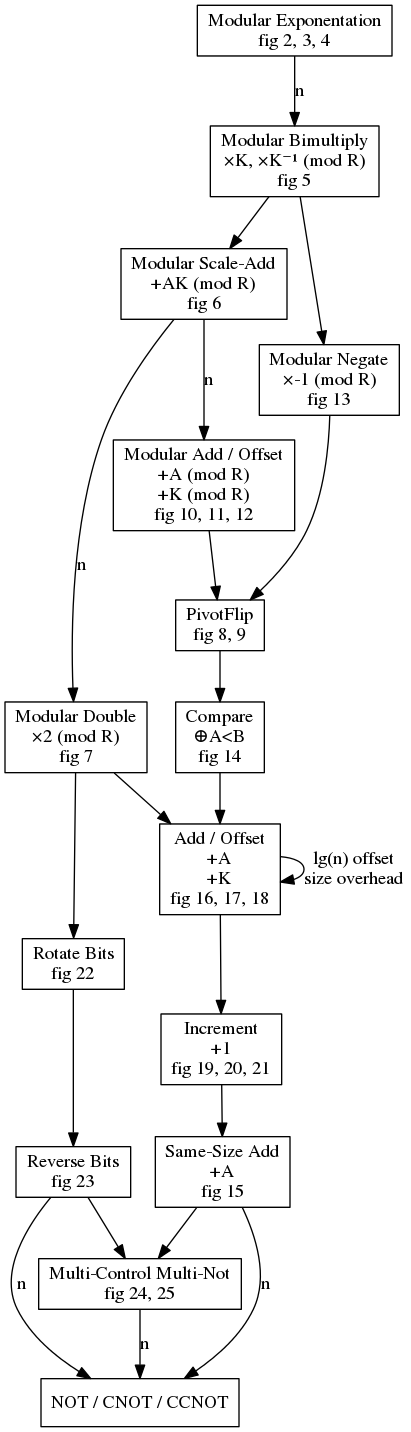}
  \caption{
    Graph of the transitive reduction of dependencies between constructions in our paper.
    Edge labels indicate which constructions use a dependency more than a constant number of times.
  }
  \label{fig:dependencies}
\end{figure}

Our reduction, from the period finding step of Shor's algorithm down to constant-sized gates, uses many circuit constructions.
\autoref{fig:dependencies} shows an overview of the various operations we will use, and the path through them that we will take.

\subsection{Period Finding for Modular Exponentiation}

\begin{figure}
  \centering
  \makebox[\linewidth]{
    \includegraphics[width=\linewidth]{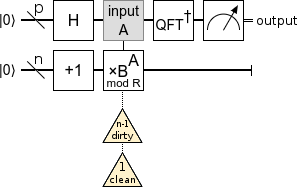}
  }
  \caption{
	High-level definition of the period finding circuit \cite{Shor1999} used by Shor's algorithm.
	$R$ is the modulus and the number to be factored, $B$ is a randomly chosen base, $n$ is the number of bits needed to store $R$, and $p \in \Theta(n)$ controls the precision of the phase estimation step.
    The triangles indicate how many ancillae are needed ``behind the scenes", by our constructions, to perform an operation.
	Recovering the period requires classical post-processing of the sampled output.
  }
  \label{fig:period-finding}
\end{figure}

A high-level view of the circuit for period finding applied to modular exponentiation, the core quantum subroutine of Shor's quantum factoring algorithm \cite{Shor1999}, is shown in \autoref{fig:period-finding}.

Before the quantum circuit is constructed, a base $B$ is chosen at random.
The randomly chosen $B$ must be co-prime to $R$.
When a $B$ that is not co-prime to $R$ is accidentally chosen, the lucky victim can instead factor $R$ by recursively factoring $R_1 = \gcd(B, R)$ and $R_2 = R / \gcd(B, R)$.

The circuit begins by preparing a uniform superposition $|\psi_0\rangle = \sqrt{2^{-n}} \sum_{k=0}^{2^n-1} |k\rangle$, then uses the $\times B^A {\pmod R}$ operation to separate that superposition into equivalence classes modulo the unknown period $l$ of this operation.
That is to say, w.l.o.g. the state ends up equal to $|\psi_{1,x}\rangle = \sqrt{l 2^{-n}} \sum_{k=0}^{\approx (2^n/l)-1} |l k + x \rangle$ for some $x$.
The circuit then applies an inverse Fourier transform to the state.
Fourier transforming a uniform signal with period $l$ produces a spectrum with peaks near $N \cdot 0/l$, $N \cdot 1/l$, $N \cdot 2/l$, ..., $N \cdot (p-1)/l$.
Shor's algorithm recovers the period $l$ by sampling values $s_i$ from this spectrum, then using a continued fractions algorithm to compute the denominator of the fraction (with denominator below $R$) that is closest to $s_i/N$.

\begin{figure*}
  \centering
  \makebox[\linewidth]{
    \includegraphics[width=\linewidth]{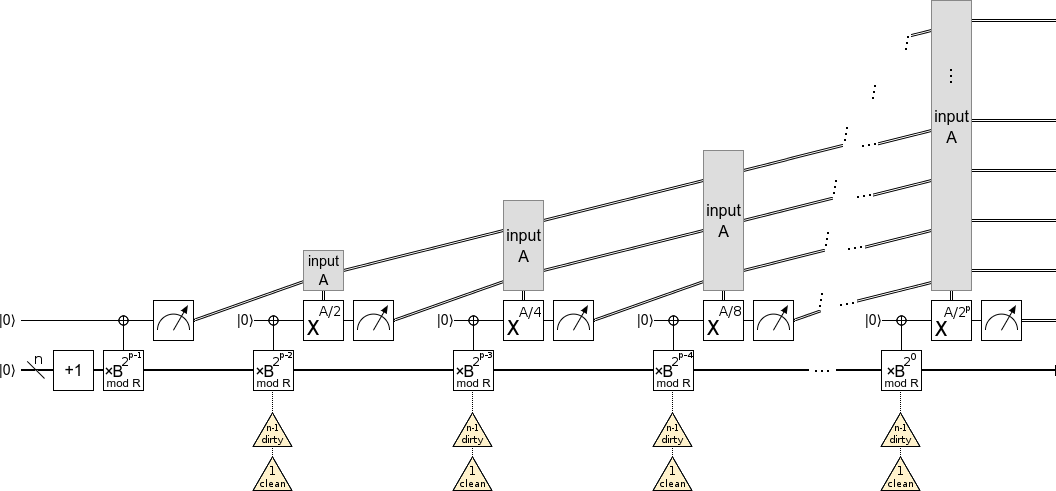}
  }
  \caption{
	Period finding with a single phase-estimation qubit \cite{beauregard2003}.
	The small oplus' ({\tiny $\oplus$}) are ``X-axis controls".
	An X-axis control is equivalent to a normal control, but with a Hadamard gate applied before and after.
	It conditions on the state $\frac{1}{\sqrt 2}|0\rangle - \frac{1}{\sqrt 2}|1\rangle$ instead of on the state $|1\rangle$.
  }
  \label{fig:period-finding-solo-phase-qubit}
\end{figure*}

Because period-finding measures all qubits immediately after performing a QFT, most of the transformed qubits can be measured earlier than shown.
In fact, each qubit can be measured so early that the next qubit needed for the QFT does not even need to be initialized yet!
Only one of the qubits in the phase-estimation register needs to be present at a time, and so the phase register can be reduced to a single repeatedly-used qubit \cite{zalka1998, mosca1999, parker2000, beauregard2003}.
\autoref{fig:period-finding-solo-phase-qubit} shows a period-finding circuit with this property.
It uses (a) controlled modular multiplication, (b) measurement, (c) X-axis rotations classically parametrized by previous measurements, and (d) qubit resets.
The only non-trivial operation is (a), the controlled modular multiplication of an $n$-qubit register.

We perform modular multiplication with modular scaled-addition operations and an ancilla register as in \cite{beauregard2003}.
However, to allow our ancilla register to be mostly dirty, we extend the operation so that it has a well defined effect on the second register: multiplying by the inverse factor.
We will refer to this combined operation as a ``bimultiply".

\begin{figure*}
  \centering
  \makebox[\linewidth]{
    \includegraphics[width=\linewidth]{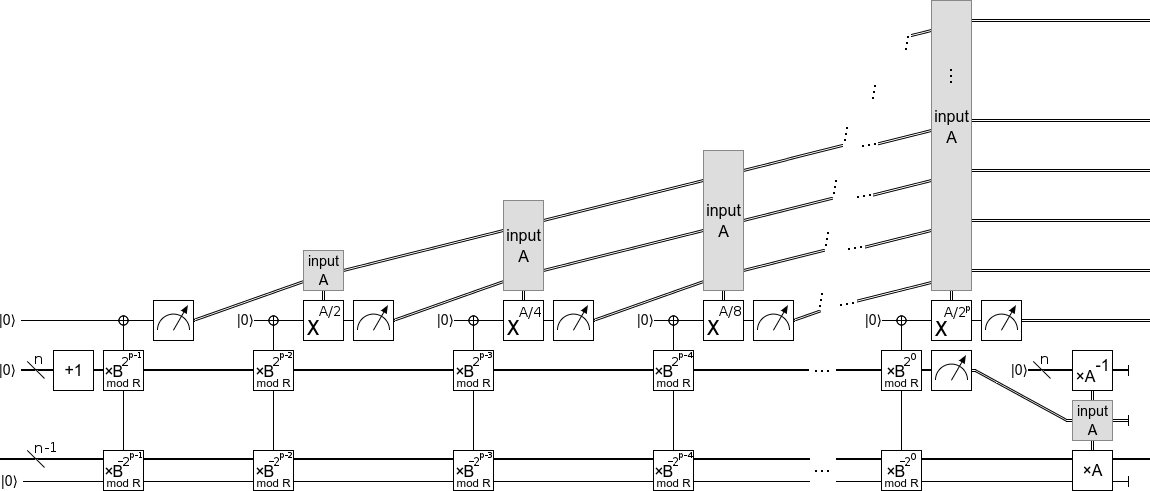}
  }
  \caption{
	Period finding with a single phase-estimation qubit and paired inverse multiplications (``bimultiplications").
	Uses $O(n^3 \lg n)$ gates, $O(n^3)$ depth, and no additional qubits beyond those shown.
  }
  \label{fig:period-finding-solo-phase-qubit-explicit-dirty-register}
\end{figure*}

The bimultiplications we perform throughout the algorithm do change the value of the ancilla register, but after the usual end of the circuit we can undo the damage.
The key insight is that the work register is initialized to $|1\rangle$ and gets multiplied by constants inverse to the ones trashing the dirty ancilla register.
Instead of discarding the work register after performing phase estimation, as is normally done, we have a use for it.
We measure the work register to recover the fixup factor needed to restore the dirty register's original value.
\autoref{fig:period-finding-solo-phase-qubit-explicit-dirty-register} shows this construction.

Because our modular circuit constructions would have undefined behavior if any register had a value equal to or larger than $R$, we need the ancilla register to contain a value less than $R$.
We ensure this by requiring that the ancilla register's MSB be $|0\rangle$ (i.e. clean) and that the registers be as small as possible (i.e. have size $n = \lceil \lg_2(R) \rceil$).

\subsection{Modular Bimultiplication}

\begin{figure}
  \centering
  \includegraphics[width=\linewidth]{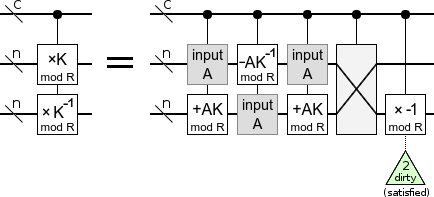}
  \caption{
    Controlled modular bimultiplication, with a constant multiplier $K$, using three modular scaled-additions and a swap.
    $K$ must have a multiplicative inverse modulo $R$.
    Uses no ancillae, $O(c + n^2 \lg n)$ gates, and $O(c + n^2)$ depth where $c$ is the number of controls and $n$ is the register size.
    The negation operation requires 2 dirty ancillae, but notice there are free wires in the other register that can be borrowed, so the overall construction requires no additional dirty ancilla (for non-trivial $n$).
  }
  \label{fig:controlled-modular-multiply}
\end{figure}

As in \cite{beauregard2003}, we perform controlled modular multiplication with a second register and modular scaled-addition operations.
However, to allow the second register to be dirty, we use an additional scale-add as well as a negation operation as in \cite{zalka2006}.
See \autoref{fig:controlled-modular-multiply} for the circuit diagram.

To show that the circuit works, suppose that the two registers start in the state $(x, y)$, working modulo $R$.
We scale-add the first register times $K$ into the second register to produce the state $(x, y+xK)$.
Then a scale-subtract times $K^{-1}$ out of the first register puts the system into the state $(x-yK^{-1}-xKK^{-1}, y+xK)$, which is just $(-yK^{-1}, y+xK)$.
Next, we cancel the $y$ term in the second register by scale-adding the first register times $K$ into it again, leaving $(-yK^{-1}, y+xK-yK^{-1} \cdot K)$ which is simply $(-yK^{-1}, xK)$.
Finally, we swap the terms and negate the second term to get $(xK, yK^{-1})$ as desired.

In the case where $K$ has no multiplicative inverse modulo $R$, this construction will not work (it would define an invalid irreversible operation).
However, that would mean $K$ is a factor of $R$; a case that can be checked for and handled classically before bothering to factor $R$ with a quantum computer.

\subsection{Modular Scaled-Addition}

\begin{figure}
  \centering
  \includegraphics[width=\linewidth]{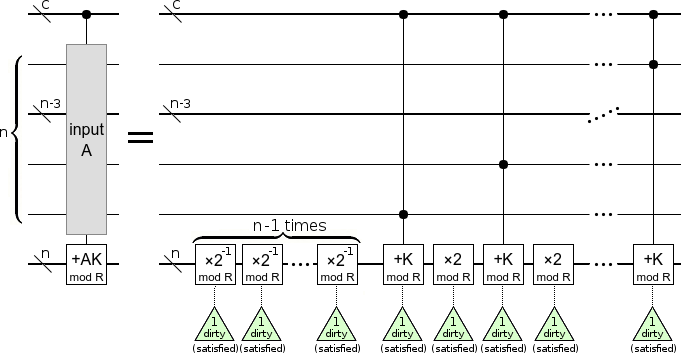}
  \caption{
    Reducing a controlled modular scale-add, with a constant multiplier $K$, into the modular equivalent of shift-and-add.
    Requires the modulus $R$ to be odd.
    Uses no ancillae, $O(c + n^2 \lg n)$ gates, and $O(c + n^2)$ depth where $c$ is the number of controls and $n$ is the register size.
  }
  \label{fig:controlled-modular-scale-accumulate}
\end{figure}

To perform modular scale-add operations, we use a shift-and-add approach similar to \cite{beauregard2003}.
We right-shift (i.e. divide by $2 {\pmod R}$) the target register $n-1$ times, then begin iteratively left-shifting (i.e. multiplying by $2 {\pmod R}$) and adding $K$ into the target.
We condition the first modular addition on the most significant bit of the input, the second addition on the next most significant bit, and so forth.
(The more significant bits go first because their effects must be hit by more left-shifts.)
See \autoref{fig:controlled-modular-scale-accumulate}.

The conditional offset and modular doubling operations need a dirty bit, but for non-trivial $n$ there are more than enough unused bits available to borrow, so the circuit as a whole doesn't require any dirty bits.

Note that the modular doublings (and halvings) require $R$ to be odd, since otherwise the operation would be irreversible.
However, given that in the intended use case $R$ is a number to be factored, it is reasonable to require callers to have factored out multiples of two beforehand.

\subsection{Modular Doubling}

To multiply a register by 2 modulo an odd $R$, we note that the permutation this operation performs is to take the values from $[(R+1)/2, R)$, the top-half of the valid range, and interleave them between the values from $[0, (R+1)/2)$, the bottom-half of the valid range.

We can perform this interleaving by moving the valid-top-half up until it starts at $2^{n-1}$, i.e. aligns with the MSB boundary.
A left-rotate of the register bits then moves the MSB to the LSB, converting the alignment into interleaving.

\begin{figure}
  \centering
  \includegraphics[width=\linewidth]{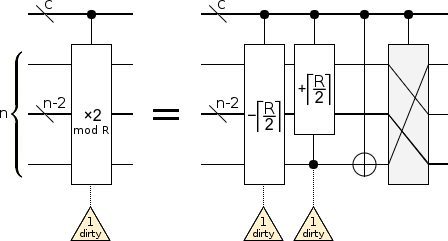}
  \caption{
    Controlled modular doubling.
    $R$ must be odd.
    Uses 1 dirty ancilla, $O(c + n \lg n)$ gates, and $O(c + n)$ depth, where $c$ is the number of controls and $n$ is the register size.
  }
  \label{fig:modular-double}
\end{figure}

Because it's expensive to perform an offset that affects values above $R/2$ while leaving values below $R/2$ alone, the circuit shown in \autoref{fig:modular-double} uses a different alignment strategy.
It starts by offset-ing the whole range down by $\lceil R/2 \rceil$, which wraps the valid-bottom-half of the input around to the top of the register's range.
This offset also aligns what was originally the valid-top-half against the bottom of the register range.
Then the circuit uses a controlled offset to move just the top half of the register (including the invalid range above $R$) up by $\lceil R/2 \rceil$, re-wrapping what was originally the valid-bottom-half past the top and around to the middle, aligning it with the MSB boundary.
The circuit then toggles the MSB, fixing the fact that, although the valid-top-half is aligned with 0 and the valid-bottom-half is aligned with $2^{n-1}$, that's exactly the reverse of what we want.
With the desired alignment achieved, the circuit interleaves the two halves with a left-rotate.

\subsection{Pivot-Flips} \label{sec:pivot-flips}

We will implement both modular negation and modular offset/addition in terms of a non-standard operation we call a ``pivot-flip".
A pivot-flip is an operation that reverses the order of states less than a given pivot value, without affecting other states.
For example, a pivot-flip with the pivot equal to 4 would swap $|0\rangle$ and $|3\rangle$, swap $|1\rangle$ and $|2\rangle$, and leave all other states untouched.

The exact permutation performed by a pivot-flip with pivot equal to $K$ is:

$$\text{PivotFlip}_K = \sum_{i=0}^{K-1} |K-i-1\rangle \langle i| + \sum_{i=K}^{N-1} |i\rangle \langle i|$$

To perform a pivot-flip efficiently, we use the fact that $x \rightarrow \lnot(x - K)$ nearly does what is required: it flips the range below $K$ but unfortunately also flips the range above-and-including $K$.
It's a ``bi-flip".

The bi-flip operation is its own inverse, so it can be ``toggle-controlled''.
If we apply two bi-flips controlled by the same control qubit, then toggling the control qubit determines whether or not a bi-flip is applied to the target register.
When the control qubit stays off, neither bi-flip fires and nothing happens to the target.
When the control qubit stays on, both bi-flips fire and they undo each other.
But if the control qubit is toggled after the first bi-flip but before the second, then exactly one of them will fire and the target register ends up bi-flipped.

Another useful property of a bi-flip is that it doesn't move any states across the pivot.
A bi-flip at $K$ preserves $x<K$ for all $x$.
This allows the toggling of the control qubit, that determines whether a toggle-controlled bi-flip will happen, to be based on a comparison of the pivot against the target register we are operating on.
Even at points where the target may or may not have been bi-flipped.

\begin{figure}
  \centering
  \includegraphics[width=\linewidth]{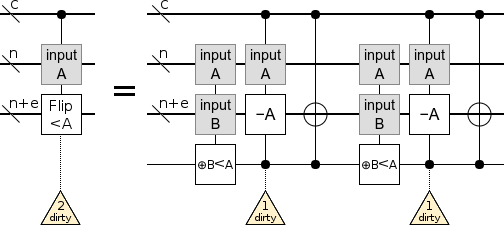}
  \caption{
    Controlled pivot-flip circuit with an enregistered pivot.
    The target register can be larger than the input register, but not smaller.
    Uses $2$ dirty ancillae, $O(c + m \lg m)$ gates, and $O(c + m)$ depth where $c$ is the number of controls, $n$ is the size of the input register, $e$ is the extra size of the target register, and $m=n+e$ is the size of the target register.
  }
  \label{fig:controlled-pivot-flip}
\end{figure}

\begin{figure}
  \centering
  \includegraphics[width=\linewidth]{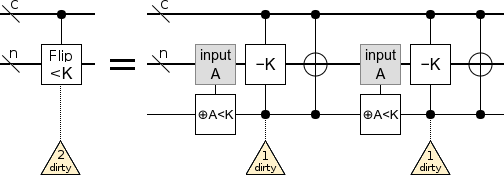}
  \caption{
    Controlled pivot-flip circuit with a constant pivot.
    Uses $2$ dirty ancillae, $O(c + n \lg n)$ gates, and $O(c + n)$ depth where $c$ is the number of controls and $n$ is the register size.
  }
  \label{fig:controlled-const-pivot-flip}
\end{figure}

See \autoref{fig:controlled-pivot-flip} and \autoref{fig:controlled-const-pivot-flip} for the circuit diagrams.
For states that are less than the pivot, the comparison against the pivot keeps toggling the ancilla and exactly one of the controlled bi-flips will fire.
For states equal to or larger than the pivot, the ancilla does not get toggled and so the bi-flips undo each other.
Therefore only the range below the pivot is flipped.

\subsection{Modular Addition / Offset}

\begin{figure}
  \centering
  \includegraphics[width=\linewidth]{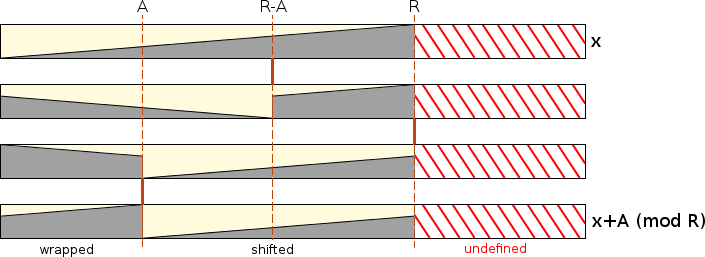}
  \caption{
     Modular addition of $A \pmod{R}$ can be done with three pivot flips.
     One at $R-A$, then one at $R$, then one at $A$.
     Requires $A \leq R$.
   }
  \label{fig:mod-add-from-pivot-flip-bars}
\end{figure}

\begin{figure}
  \centering
  \includegraphics[width=\linewidth]{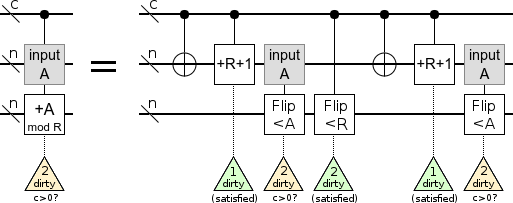}
  \caption{
    Controlled modular addition construction based on pivot-flips.
    The arithmetic being performed on the input register temporarily transitions it from storing $x$ to $R-x$ for the first pivot flip.
    Because $R$ is a compile-time constant, $R+1$ is also a compile-time constant and the $+R+1$ operation is a normal offset operation.
    All arithmetic is two's complement, i.e. modulo $2^n$.
    Uses $2-c$ dirty ancillae, $O(c + n \lg n)$ gates, and $O(c + n)$ depth where $c$ is the number of controls and $n$ is the register size.
  }
  \label{fig:controlled-modular-add}
\end{figure}

As shown in \autoref{fig:mod-add-from-pivot-flip-bars}, a modular addition can be implemented by three pivot-flips.
To add $K$ into a register modulo $R$, perform pivot-flips with the pivot at $R-K$, then $R$, then $K$.
See \autoref{fig:controlled-modular-add} for the circuit.
(An optimization we don't show is that, because the values above $R$ don't matter, the pivot-flip at $R$ can be replaced by a bi-flip.)

\begin{figure}
  \centering
  \includegraphics[width=\linewidth]{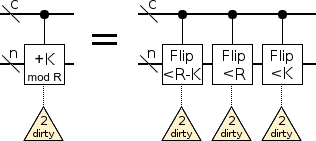}
  \caption{
    Controlled modular offset construction based on pivot-flips.
    Uses $2$ dirty ancillae, $O(c + n \lg n)$ gates, and $O(c + n)$ depth where $c$ is the number of controls and $n$ is the register size.
  }
  \label{fig:controlled-modular-offset}
\end{figure}

Interestingly, controlled modular addition can borrow its own controls as dirty bits.
For modular offset (i.e. adding a compile-time constant into a register), the two dirty bits are required whether or not the operation is controlled.
See \autoref{fig:controlled-modular-offset}.

\subsection{Modular Negation}

\begin{figure}
  \centering
  \includegraphics[width=\linewidth]{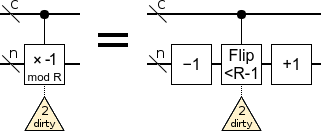}
  \caption{
    Controlled modular negation.
    Uses $2$ dirty ancillae, $O(c + n \lg n)$ gates, and $O(c + n)$ depth where $c$ is the number of controls and $n$ is the register size.
  }
  \label{fig:negate-mod}
\end{figure}

To negate a number mod $R$, we need to reverse the order of the states $|1\rangle$ to $|R-1\rangle$.
We do so by temporarily moving $|0\rangle$ out the way with a decrement, pivot-flipping at $R-1$, then undoing the decrement.
See \autoref{fig:negate-mod}.

\subsection{Comparison}

\begin{figure}
  \centering
  \includegraphics[width=\linewidth]{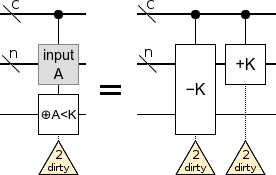}
  \caption{
    Controlled comparison.
    Uses $2$ dirty ancillae, $O(c + n \lg n)$ gates, and $O(c + n)$ depth where $c$ is the number of controls and $n$ is the register size.
  }
  \label{fig:compare}
\end{figure}

Comparison operations toggle a target bit based on the relationship between two input registers.
We implement comparisons as in \cite{takahashi2005}, using an addition followed by a slightly smaller subtraction that clears all changes except the overflow signal into the target bit.
See \autoref{fig:compare}.

For comparisons against another register this approach uses $1$ dirty ancilla (if controlled, otherwise no ancillae), $O(c + n)$ gates, and $O(c + n)$ depth.
For comparisons against a constant, the number of gates increases to $O(c + n \lg n)$ and an extra dirty ancilla is required.

When $n$ dirty qubits are available, the overflow-predicting construction from \cite{haner2016} with better constant factors can be used instead.

\subsection{Addition / Offset}

\begin{figure}
  \centering
  \includegraphics[width=\linewidth]{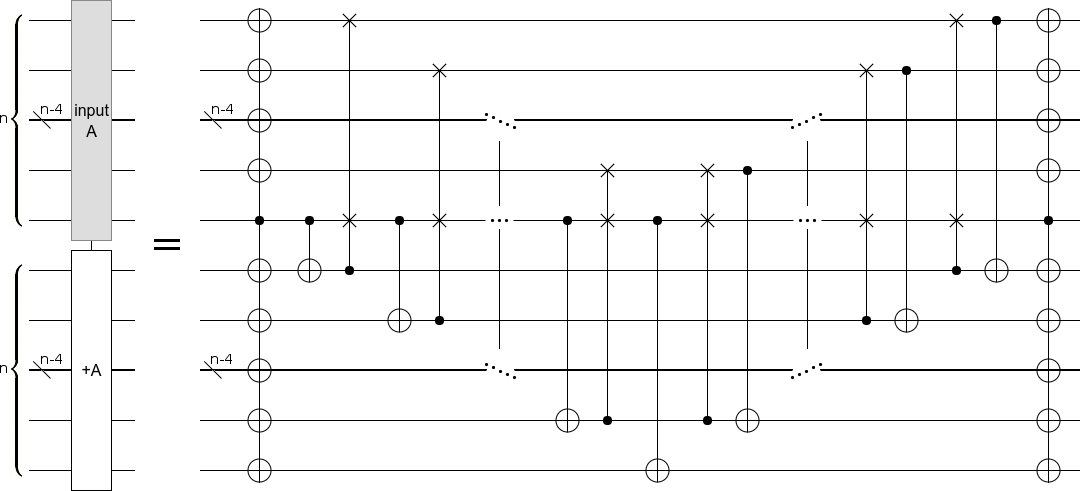}
  \caption{ Adder for input and target registers of the same size.
  Requires no ancillae, and uses $O(n)$ gates and depth.
  Based on \cite{van2004, takahashi2005}.}
  \label{fig:inlineadder}
\end{figure}

\cite{takahashi2005} provides a reversible adder circuit that uses $O(n)$ gates, $O(n)$ depth, and no ancillae.
We show an equivalent circuit in \autoref{fig:inlineadder}.

Because the construction in \autoref{fig:inlineadder} uses the input register as workspace, it doesn't work when the input is a compile-time constant or when the target register is larger than the input register.
(When the target register is smaller, we simply ignore the high bits of the input due to overflow wraparound.)
This is a problem for us, because some of our constructions add constants into register (e.g. modular scale-add), and some others add inputs into larger target registers (e.g. comparison).

\begin{figure}
  \centering
  \includegraphics[width=\linewidth]{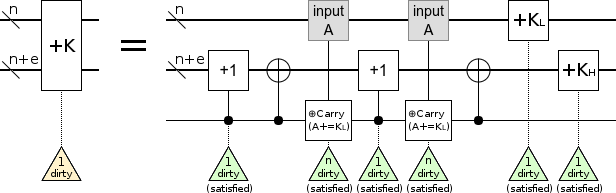}
  \caption{
      Offset circuit from \cite{haner2016}.
      $e$ is either 0 or 1.
      Uses $1$ dirty ancilla, $O(n \lg n)$ gates, and $O(n)$ depth (by overlapping the recursive cases).
  }
  \label{fig:offset}
\end{figure}

When the value to be added into a register is a compile-time constant (i.e. when applying an offset gate), we use the offset construction from \cite{haner2016}.
Their offset circuit, shown in \autoref{fig:offset}, uses $O(n \lg n)$ gates, $O(n)$ depth, and one dirty ancilla.

When the target register is larger than the input register, the addition circuit in \autoref{fig:inlineadder} can be modified without increasing the asymptotic cost.
We do this by removing the surrounding CNOTs and adding three increment/decrement operations to the circuit, as follows.
(Note that, to avoid cyclic dependencies, the increment and decrement constructions described in the next subsection will only use the same-register-size adder.)

\begin{figure}
  \centering
  \includegraphics[width=\linewidth]{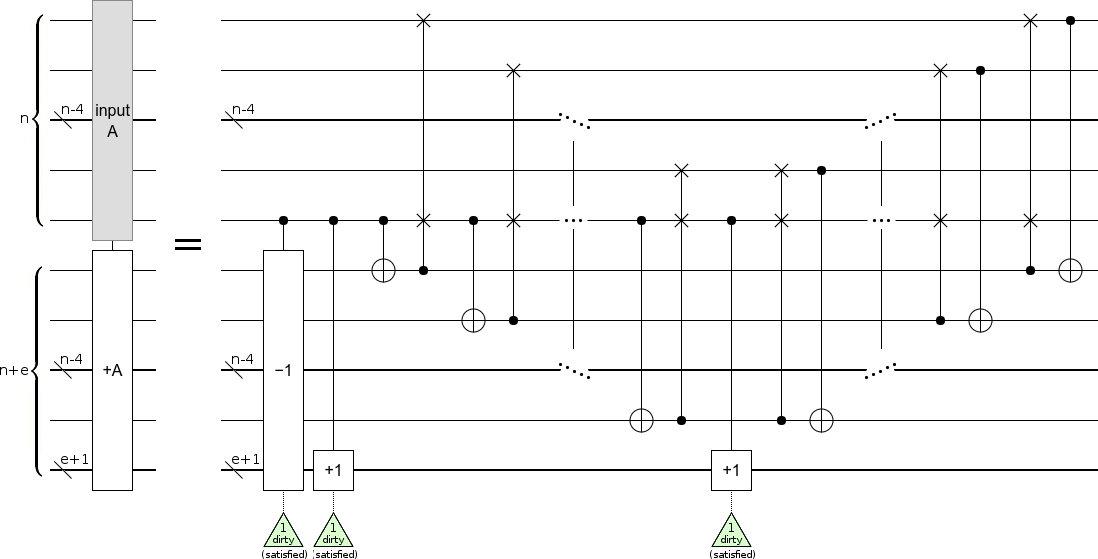}
  \caption{
      Adder with target larger than source, using no ancillae.
      Uses $O(n)$ gates and depth.
      The increment and decrement gates all have at least one free wire to borrow as a dirty ancilla.
  }
  \label{fig:inline-adder-into-large}
\end{figure}

First, because there are many target bits not reached by the time we have swept through the entire input register, carry signals aren't reaching the high bit of the target register.
We fix this by replacing the innermost CNOT, the carry-propagating CNOT, with a controlled increment.
Second and third, because we're using the MSB of the input register as the carry signal, it causes an increment at the LSB instead of at the correct position.
We undo the LSB increment with a controlled decrement, and use a controlled increment to apply the MSB's effect where it should have actually gone.
See the circuit diagram in \autoref{fig:inline-adder-into-large}.

To control addition gates, and offset gates, we use ``commutator controlling": finding operations $G$ and $H$ such that their group-theory commutator $[G, H] = G \cdot H \cdot G^\dagger \cdot H^\dagger$ equals a desired operation $U$ and either $G$ or $H$ can be efficiently controlled.
In particular, we focus on the case where $G^2 = U$ and $H$ satisfies $H \cdot G \cdot H^\dagger = G^\dagger$ (i.e. framing $G$ with $H$ inverts the effect of $G$).

\begin{figure}
  \centering
  \includegraphics[width=\linewidth]{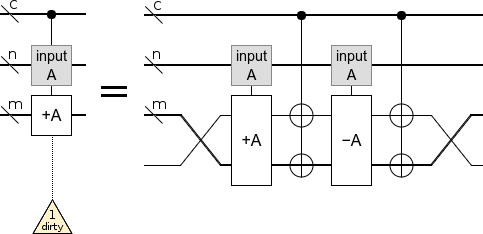}
  \caption{
    Reducing controlled addition to uncontrolled addition.
    Uses $O(c + m)$ gates, $O(c + m)$ depth, and a dirty ancilla.
    When applied to offset gates, uses $O(c + m \lg m)$ gates, $O(m + c)$ depth, and two dirty ancillae.
    $c$ is the number of controls, $n$ is the size of the input, and $m > n$ is the size of the target register.
  }
  \label{fig:controlled-addition}
\end{figure}

In the case of addition, we have the property $\lnot (\lnot x + K) = \lnot (-x - 1 + K) = -(-x - 1 + K) - 1 = x-K$.
So a valid inverting operation $H$ for addition is a NOT gate applied to every target wire; $H=X^{\otimes n}$.
If we then consider $G=\text{S}_{\text{-} K}$, where $S$ is an offset-by-subscript operation, we find that $[G, H] = [\text{S}_{-K}, X^{\otimes n}] = \text{S}_{2K}$.
We can control this constructed $\text{S}_{2K}$ operation by adding controls only to the multi-not operation $H$, which is cheap to do.
To add $K$ instead of $2K$, we temporarily prepend a dirty LSB onto the target register.
We show this construction in \autoref{fig:controlled-addition}.

\subsection{Increment}

\begin{figure}
  \centering
  \includegraphics[width=\linewidth]{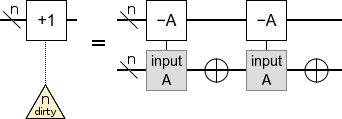}
  \caption{ Subtracting $x$ and $-x-1$ from a register increments it. Requires $O(n)$ depth and size, and $n$ dirty ancillae.}
  \label{fig:increment-many-dirty}
\end{figure}

A register can be incremented by subtracting both $x$ and $\neg x = -x-1$ from it, for any $x$.
When $n$ dirty bits are available, $x$ can come from a register defined by those $n$ arbitrary bits, as shown in \autoref{fig:increment-many-dirty}.

\begin{figure}
  \centering
  \includegraphics[width=\linewidth]{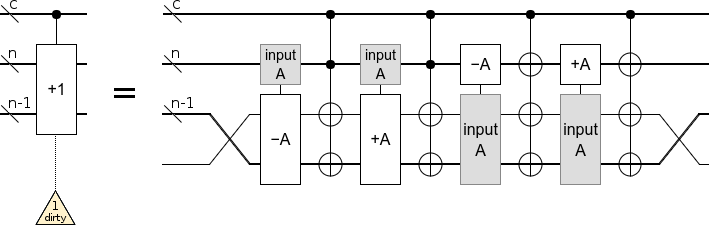}
  \caption{
    Odd-sized controlled increment.
    For the even-sized case, separate the LSB from the rest of the register, increment the rest of the register using the LSB as an extra control, then toggle the LSB.
    Uses $O(c+n)$ gates, $O(c+n)$ depth, and 1 dirty ancilla.
  }
  \label{fig:controlled-increment-odd}
\end{figure}

To improve from $n$ dirty bits to the single dirty bit used by the circuit in \autoref{fig:controlled-increment-odd}, we break the register into two halves.
A high-half that is incremented only if all of the low bits are on, and a low-half that is unconditionally incremented.

If the increment is uncontrolled, the low-half can be incremented with the double-subtraction trick by borrowing the high-half.
When there are controls, we instead increment the low-half using a commutator-control construction where one operation subtracts the ancilla register out of the target and the other applies a NOT gate to every input wire and target wire.
This works because $\lnot (\lnot T + \lnot K) - K = \lnot (-T - 1 - K - 1) - K = (T+2+K) - 1 - K = T + 1$.

The high-half is trickier to deal with.
We want to borrow the low-half for the double-subtraction trick, but the low-half is being used as a control and so can't be borrowed.
To work around not being able to operate on the borrowed low-half bits while using them as a control, we use more commutator control tricks and some knowledge of what state the low-half must be in if operations it is controlling are firing.

We add and subtract the low-half out of the high-half, but frame the addition with NOT gates controlled by all bits in the low half.
When any of the bits in the low-half are off, the NOT gates don't fire and the addition and subtraction will cancel each other.
When all of the bits in the low-half are on, i.e. when the low-half is storing the two's complement representation of -1, the NOT gates do fire.
This inverts the addition into a subtraction, and the low-half (which is storing -1) is subtracted out of the high-half twice.
Therefore the high-half was incremented by 2.
To halve the +2 into a +1, we prepend a dirty LSB onto the target register.

Recall that, earlier in the paper, we used increments and decrements to implement addition where the target was larger than the input.
To avoid an expensive cyclic dependency in our increment construction, the additions we use must not be larger-target additions.
(Also, when the two registers are not the same size, they define arithmetic modulo different powers of 2.
This breaks properties such as $T + \lnot K = T - K - 1$.)

Because we can only perform additions and subtractions with input and target registers of the same size, the construction described so far in this section only works for odd-sized registers.
To handle even-sized registers, we decompose the increment into a controlled-increment and a NOT gate.
We use the LSB as a control determining whether the (odd-sized) rest of the register is incremented, and then toggle the LSB.

\begin{figure}
  \centering
  \includegraphics[width=\linewidth]{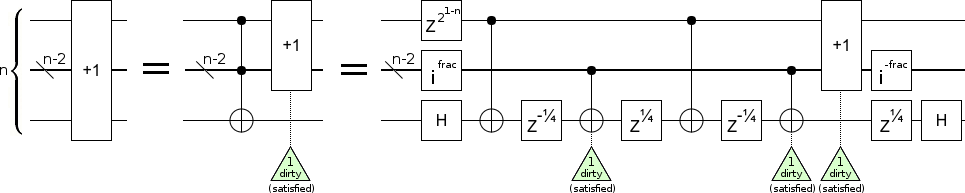}
  \caption{ Bootstrapping a dirty ancilla out of an increment gate using quantum operations.
  The $i^{\text{frac}}$ gate is a ``phase gradient" operation that phases each computational basis state $|v\rangle$ by an amount proportional to $v/2^d$, where $d$ is the size of the register.
  In this case each state is phased by $e^{i \frac{\pi}{2} v/2^d}$.
  The phase gradient is implemented by a column of $Z^{2^{-k}}$ gates.}
  \label{fig:bootstrap-ancilla}
\end{figure}

\subsection{Bit Swaps, Rotations, and Reversals}

\begin{figure}
  \centering
  \includegraphics[width=\linewidth]{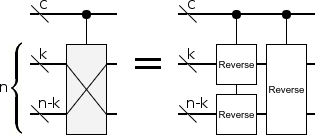}
  \caption{
    A controlled bit rotation / bit swap is three controlled bit-reverses.
    Uses no ancillae, $O(c + n)$ gates, and $O(c + n)$ depth.
  }
  \label{fig:bit-rotate}
\end{figure}

\begin{figure}
  \centering
  \includegraphics[width=\linewidth]{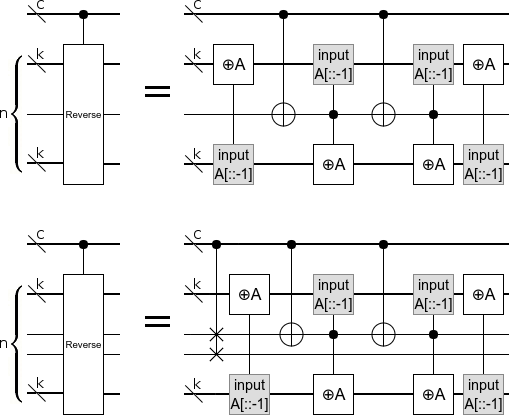}
  \caption{
    Inline controlled bit order reversal on an odd-sized register of size $n=2k+1$, and on an even-sized register of size $n=2k+2$.
    Each XOR operation is a series of independent CNOTs (note that the inputs have opposite endian-ness to the outputs).
    Uses no ancillae, $O(c + n)$ gates, and $O(c + n)$ depth.
  }
  \label{fig:bit-reverse}
\end{figure}

Bit permuting operations can usually be emulated by re-labelling qubits, so they are easy to overlook in circuits.
But some of our circuit diagrams have used controlled bit rotations which require actual gates.
We provide the relevant constructions in \autoref{fig:bit-rotate} and \autoref{fig:bit-reverse}.

\subsection{Multi-Nots}

Several of our constructions have used CNOTs applied to wire bundles, with many controls and many targets.
A naive approach to implementing these operations would be to apply a separate NOT, each controlled by every control, to every target.
But this would use $O(n \cdot c)$ gates, which is not linear in the number of controls.

To avoid paying the overhead of $c$ controls for every target, we toggle {\em one} target conditioned on all of the controls.
We then use toggle-controlling to spread the toggling effect to all of the other targets.

\begin{figure}
  \centering
  \includegraphics[width=\linewidth]{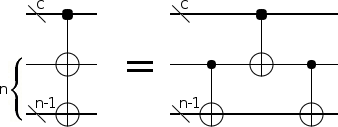}
  \caption{
    Reducing a many-control many-target CNOT into one many-control single-target CNOT and many single-target single-control CNOTs.
    Uses no ancillae, $O(c + n)$ gates, and $O(c + n)$ depth where $n$ is the number of targets and $c$ is the number of controls.
    The depth can be reduced to $O(c + \lg n)$ by spreading the toggling effect more intelligently.
  }
  \label{fig:multi-not}
\end{figure}

\begin{figure}
  \centering
  \includegraphics[width=\linewidth]{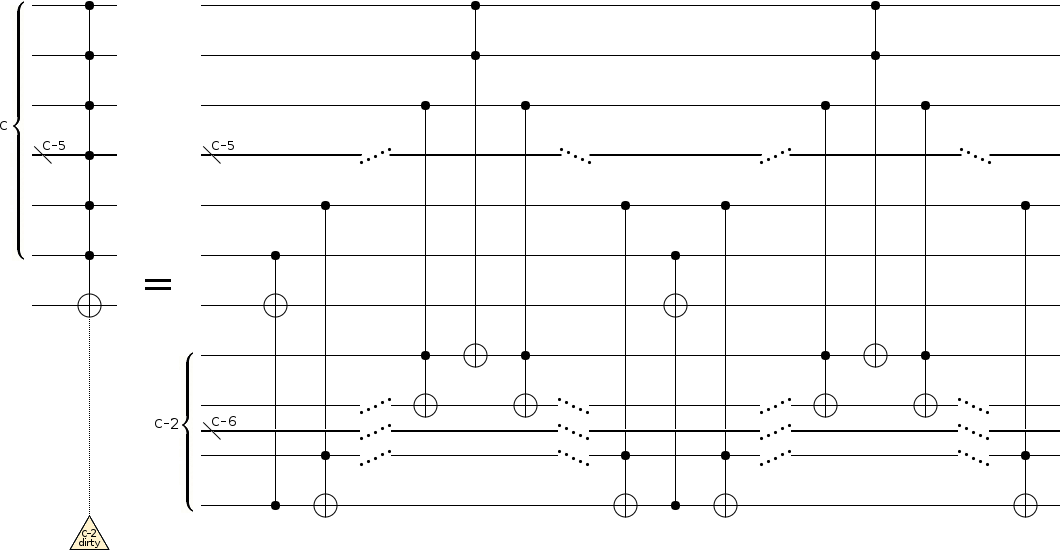}
  \caption{
    When $c-2$ dirty ancillae are available, a controlled-not with $c$ controls can be reduced into $4c - 8$ Toffoli gates \cite{barenco1995}.
    It is possible to reduce the number of ancilla needed to just one, by using twice as many Toffoli gates, but that construction isn't needed in the context of period finding because there are always enough unused qubits to borrow.
  }
  \label{fig:cnot-reduction}
\end{figure}

To efficiently reduce the single remaining CNOT with $c$ controls into constant-sized Toffoli gates, we take advantage of the many available dirty ancilla.
See \autoref{fig:multi-not} and \autoref{fig:cnot-reduction}.

\section{Overview and Improvements} \label{sec:costs}

\begin{figure*}
  \centering
  \makebox[\linewidth]{
    \begin{tabular}{ l | c c c c c l }
       & Year & Depth & Gates & Clean Qubits & Total Qubits \\
      \hline
      Shor \cite{Shor1999} & 1994 & $\Theta(n M(n))$ & $\Theta(n M(n))$ & $\Theta(n)$ & $\Theta(n)$ \\
      Beckman et al. \cite{beckman1996} & 1996 & $\Theta(n^3)$ & $\Theta(n^3)$ & $5n + 1$ & $5n + 1$ \\
      Veldral et al. \cite{vedral1996} & 1996 & $\Theta(n^3)$ & $\Theta(n^3)$ & $4n + 3$ & $4n + 3$ \\
      Beauregard \cite{beauregard2003} & 2003 & $\Theta(n^3 \lg \frac{1}{\varepsilon})$ & $\Theta(n^3 \lg \frac{n}{\varepsilon} \lg \frac{1}{\varepsilon})$ & $2n+3$ & $2n+3$ \\
      Takahashi et al. \cite{takahashi2006} & 2006 & $\Theta(n^3 \lg \frac{1}{\varepsilon})$ & $\Theta(n^3 \lg \frac{n}{\varepsilon} \lg \frac{1}{\varepsilon})$ & $2n+2$ & $2n+2$ \\
      Zalka \cite{zalka2006} & 2006 & $\Theta(n^3 \lg \frac{1}{\varepsilon})$ & $\Theta(n^3 \lg \frac{n}{\varepsilon} \lg \frac{1}{\varepsilon})$ & $1.5n+O(1)$ & $1.5n+O(1)$ \\
      H\"{a}ner et al. \cite{haner2016} & 2016 & $\Theta(n^3)$ & $\Theta(n^3 \lg n)$ & $2n+2$ & $2n+2$ \\
      (ours) & 2017 & $\Theta(n^3)$ & $\Theta(n^3 \lg n)$ & $n+2$ & $2n+1$ \\
    \end{tabular}
  }
  \caption{
    Space-efficient constructions of Shor's algorithm over time.
    The table only includes papers that presented explicit circuit constructions and improved the minimum number of qubits required to perform the algorithm (or else the time complexity at a slightly larger qubit count).
    $M(n)$ is the classical time-complexity of multiplication, which is known to be asymptotically at most $n \cdot (\lg n) \cdot 2^{O(\lg^* n)}$ \cite{furer2007}.
    $\varepsilon$ is the maximum error when synthesizing the circuit out of a fixed set of universal gates, which is asymptotically relevant for algorithms that use Draper addition \cite{draper2000} (i.e. applying phase gradients in frequency space) instead of Toffoli-based addition/offset constructions.
  }
  \label{fig:table-over-time}
\end{figure*}

Recall that \autoref{fig:dependencies} shows a dependency graph of the constructions discussed in this paper.
Our asymptotic costs are dominated by performing $O(n)$ modular multiplications, each of which uses $O(n)$ modular additions and offsets, each of which uses $O(n \lg n)$ constant-sized classical gates \cite{haner2016} and $O(n)$ depth.
The total cost of the period-finding step in Shor's algorithm, using our construction, is $O(n^3 \lg n)$ gates and $O(n^3)$ depth.

In \autoref{fig:table-over-time} we show how the number of qubits needed for Shor's algorithm has improved over time.
Our main improvements over previous arithmetic constructions are 1) the use of pivot-flips for modular addition, 2) the use of dirty bimultiplication for modular multiplication, and 3) the $O(n)$ incrementer requiring only a single dirty ancilla.

Previous modular addition constructions worked by temporarily storing an is-wraparound-needed comparison in a clean ancilla \cite{takahashi2006, haner2016}.
Pivot flips also require ancillae, but the ancillae can be dirty and, in the context of Shor's algorithm, there are always qubits available to borrow whenever a pivot-flip is needed.
This improvement ends up saving a qubit, reducing the total number of qubits we would have needed from $2n+2$ as in \cite{haner2016} to $2n+1$.

Previous modular multiplication constructions did not work with a dirty ancilla register.
They either required a clean ancilla register because they leaked the ancilla register into the work register \cite{haner2016}, or else a trashable ancilla register because there was no way to undo the damage being done by the multiplications \cite{zalka2006}.
The leakage problem is fixed by adding an extra scale-add operation and a negation operation \cite{zalka2006}.
We fixed the trashing problem by measuring the work register at the end of the circuit, and using its value to drive a clean-up multiplication that restores the ancilla register.
This improvement allows $n-1$ of the qubits in the ancilla register to be dirty.
The MSB has to stay clean to ensure the register's value is less than the modulus being factored.

Previous published incrementers (not counting an unpublished version of our construction being cited by \cite{haner2016}) required either $O(n^2)$ gates or $\omega(1)$ ancillae \cite{draper2000, barenco1995}.
Our classical incrementer construction uses $O(n)$ gates and a single dirty ancilla.

Note that, for classical reversible computation, 1 dirty ancilla is the minimum possible for an incrementer.
An increment operation on $n$ bits is equivalent to performing a state-permutation that uses $2^n-1$ swaps to sweep the state $|2^n-1\rangle$ from the top of the state-space to the bottom of the state-space one step at a time.
Note that $2^n-1$ is odd for non-trivial $n$, and therefore the parity of the state permutation performed by an increment operation is odd.
However, the parity of the permutation performed by any classical gate that doesn't cover the entire circuit is even.
Since odd-parity permutations can't be implemented by composing even-parity permutations, it is impossible to compose smaller operations into an increment operation that covers every bit.
An uncovered bit (i.e. a dirty ancilla) must be present.

When quantum operations are available, the parity barrier can be bypassed and incrementing can be performed without any ancilla.
The asymptotic cost is slightly worse due to the cost of synthesizing tiny rotations, and the constant factors hidden by the asymptotic notation are also worse.
See \autoref{fig:bootstrap-ancilla} for the construction.

We used \href{https://github.com/Strilanc/Quirk}{Quirk} \cite{quirk2016} to explore \cite{victor2013}, check, and refine individual constructions.
The full construction down to Toffoli gates was tested in \href{https://github.com/ProjectQ-Framework/ProjectQ}{ProjectQ} \cite{projq2016}.
The python source code for our test implementation, and an issue tracker for submitting errata, can be found online at \href{https://github.com/Strilanc/PaperImpl-2017-DirtyPeriodFinding}{https://github.com/Strilanc/PaperImpl-2017-DirtyPeriodFinding}.

Although the asymptotic costs of our constructions match previous work, the constant factors are significantly worse.
For example, our equivalent of a controlled 32-qubit multiplication uses roughly 1.3 million Toffoli gates (the specific number depends on the factor to multiply by, the modulus, whether or not a number of optimizations are used, and whether or not those optimizations apply to a given case).
This cost is over an order of magnitude worse than previous work \cite{haner2016}.
The reason for the increased cost is that using dirty ancillae results in more uncomputation work, in repeating some operations twice conditionally to do them once unconditionally, and in less efficient constructions in general.
These inefficiencies stack multiplicatively when one construction uses another as a subroutine.

\section{Future Work and Conclusion} \label{sec:conclusion}

In this paper we described how to perform various arithmetic operations using only dirty ancilla.
We also described various techniques we used to find the constructions: pivot flipping, toggle controlling, commutator controlling, and exploration by directly manipulating circuit diagrams.
We showed how these constructions reduce the number of clean qubits required to perform Shor's algorithm from $1.5n+O(1)$ to $n+2$.

Of the $2n+1$ qubits our period-finding construction requires, $1$ is used for the phase estimation qubit, $n$ are used to store the work register, and $n$ are ancillae used to implement modular multiplication in terms of scaled modular addition.
Except for multiplication, all our arithmetic constructions use two or fewer dirty ancillae.
Since we reduce modular scale-addition into operations that have a $\Theta(n)$ surplus of unused qubits, in context there are always more than enough dirty ancillae available to implement the simpler arithmetic inline.
Because there is slack in the simpler arithmetic constructions, improving the overall number of qubits used by Shor's algorithm requires only that we improve the first step of reducing modular multiplication into some other operation.

One way we could reduce modular multiplication into better operations is by knowing the factorization $p \cdot q$ of the modulus $R$, with $p$ and $q$ each having size $\approx n/2$.
The Chinese remainder theorem guarantees that $x \mod R$ could be uniquely represented as the pair of half-sized values $(x \mod p, x \mod q)$.
Multiplications of the half-sized values could use and reuse the same half-sized ancilla register, for a total of three half-sized registers (one for $x \mod p$, one for $x \mod q$, and one hopefully-dirty ancilla register).
Despite the large reduction in the number of ancillae available to the underlying arithmetic operations, each operation would still have more than enough ancillae and contribute no additional ancillae to the overall circuit.
(Note: because the initial value $x_0 = 1$ is trivial, and the final value is measured or discarded, it's not necessary to implement circuits that translate between the Chinese-remainder representation and the usual 2s-complement representation.)
Of course, since we use Shor's algorithm to compute the factorization of $R$, using the factorization of $R$ to optimize Shor's algorithm would be paradoxical.
The $1.5n$ achieved in \cite{zalka2006} was done via this kind of method, except that instead of factoring $R$ Zalka factored the value to multiply by and the factoring was done modulo $R$ (by stopping a generalized gcd at a midway point) instead of in $\mathbb{Z}$.

A second way to reduce modular multiplication into (slightly) smaller operations could be with commutator control.
Note that the modular Fourier transform inverts the effect of modular multiplication: $QFT_{R} \cdot (\times K \,\text{mod}\, R) \cdot QFT_R^{-1} = (\times K^{-1} \,\text{mod}\, R)$.
This has two useful high-level effects.
First, by applying the bimultiplication gate used in this paper twice, but inverting the effect on the ancilla register for the second application, we can build a proper modular multiplication gate that only affects a single register.
Second, we can use commutator control to move controls from the modular multiplication gate onto modular Fourier transform gates.
Assuming controlled modular QFTs can be performed with fewer ancillae than controlled modular multiplication, we could save a qubit when performing the modular multiplication by borrowing the control qubit.

Possible future improvements aside, in this paper we showed that the number of clean qubits sufficient to perform Shor's algorithm is no more than $n + O(1)$.
We consider this to be a step towards a construction for Shor's algorithm that uses only $n + O(1)$ qubits total.
We do note, however, that these improvements are of more theoretical than practical interest.
In practice, on error corrected quantum computers, the important cost to optimize is the number of T gates.
In that sense, the main {\em practical} contribution of this paper is the improved increment circuit and the demonstration of constructing circuits under tight space constraints.

\section{Acknowledgements}

We thank Matthew Neeley, Dave Bacon, and Austin Fowler for comments on earlier versions of this paper, from which it greatly benefited.

\bibliographystyle{plain}
\bibliography{citations}

\end{document}